\documentclass{appolb}
\usepackage{epsfig}
\begin{document}

%------------------------BEGINNING OF CERN TITLE PAGE----------------------

\thispagestyle{empty}

\begin{flushright}
CERN-PH-TH/2005-264\\
hep-ph/0512271
\end{flushright}

\vspace{1.0truecm}
\begin{center}
\boldmath
\large\bf CP Violation in the $B$ System: Status and Perspectives
\unboldmath
\end{center}

\vspace{0.9truecm}
\begin{center}
Robert Fleischer\\[0.1cm]
{\sl CERN, Department of Physics, Theory Division\\
CH-1211 Geneva 23, Switzerland}
\end{center}

\vspace{0.9truecm}

\begin{center}
{\bf Abstract}
\end{center}

{\small
\vspace{0.2cm}\noindent
In this decade, the exploration of CP violation is governed by the 
$B$-meson system, where non-leptonic decays play an outstanding
r\^ole. To set the stage, we have first a look at the main strategies to circumvent 
the calculation of the relevant hadronic matrix elements in these studies, 
and discuss popular avenues for physics beyond the Standard Model
to enter the roadmap of quark-flavour physics. We are then well prepared
to analyze puzzles in the data for $B\to\phi K$ and $B\to \pi K$ decays, and to 
discuss $b\to d$ penguin processes, which represent a new territory for the 
$B$-factory studies. Finally, we turn to the physics potential of the $B_s$-meson 
system, which can be fully exploited at the LHC.
}

\vspace{0.9truecm}

\begin{center}
{\sl Invited talk at the GUSTAVOFEST\\
Symposium in Honour of Gustavo C. Branco\\ 
``CP Violation and the Flavour Puzzle"\\
Lisbon, Portugal, 19--20 July 2005\\
To appear in the Proceedings (Acta Physica Polonica)}
\end{center}

\vfill
\noindent
CERN-PH-TH/2005-264\\
December 2005

\newpage
\thispagestyle{empty}
\vbox{}
\newpage
 
\setcounter{page}{1}

%------------------------END OF CERN TITLE PAGE------------------------------

\title{CP VIOLATION IN THE B SYSTEM: STATUS AND PERSPECTIVES%
\thanks{Invited talk at the GUSTAVOFEST -- Symposium in Honour of 
Gustavo C. Branco: ``CP Violation and the Flavour Puzzle", 
Lisbon, Portugal, 19--20 July 2005.}%
}
\author{Robert Fleischer
\address{CERN, Department of Physics, Theory Division, CH-1211 Geneva 23, 
Switzerland}
}
\maketitle
\begin{abstract}
In this decade, the exploration of CP violation is governed by the 
$B$-meson system, where non-leptonic decays play an outstanding
r\^ole. To set the stage, we have first a look at the main strategies to circumvent 
the calculation of the relevant hadronic matrix elements in these studies, 
and discuss popular avenues for physics beyond the Standard Model
to enter the roadmap of quark-flavour physics. We are then well prepared
to analyze puzzles in the data for $B\to\phi K$ and $B\to \pi K$ decays, and to 
discuss $b\to d$ penguin processes, which represent a new territory for the 
$B$-factory studies. Finally, we turn to the physics potential of the $B_s$-meson 
system, which can be fully exploited at the LHC.
\end{abstract}

\section{Setting the Stage}\label{sec:intro}
\subsection{The Flavour-Physics Landscape in a Nutshell}
As discussed in detail in the {\it GustavoBook} \cite{G-Book}, CP violation 
is a particularly interesting phenomenon in particle physics, which offers a 
powerful tool to explore the flavour sector of the Standard Model (SM) and 
to search for signals of  ``new physics" (NP). After a long and exciting 
history of $K$-decay studies, the stage is now governed by the decays of
$B^+$ and $B^0_d$ mesons. Thanks to the efforts at the $e^+e^-$ $B$ factories 
with their detectors BaBar (SLAC) and Belle (KEK), CP violation is now 
also in the $B$-meson system well established, and several strategies to test the 
flavour structure of the SM, which is governed by the
Cabibbo--Kobayashi--Maskawa (CKM) matrix \cite{CKM}, can now be confronted -- 
for the first time -- with experimental data. These experiments have already collected 
${\cal O}(10^8)$ $B\bar B$ pairs. Further valuable insights can be obtained through 
the studies of the $B^0_s$-meson system. After first results from the LEP 
experiments (CERN) and SLD (SLAC) as well as from the Tevatron, the 
physics potential of $B^0_s$ decays can be fully exploited at the Large Hadron 
Collider (LHC) at CERN, in particular by the LHCb experiment \cite{Nakada}.
Moreover, there are also plans for a ``super-$B$ factory", with a significant 
increase of luminosity relative to the currently operating $e^+e^-$ colliders. 
As far as the kaon system is concerned, the future lies in particular on the 
investigation of the very rare decays $K^+\to\pi^+\nu\bar\nu$ and 
$K_{\rm L}\to\pi^0\nu\bar\nu$, which are very clean from the theoretical point 
of view, but unfortunately hard to measure; there is a new proposal to measure 
the former channel at the CERN SPS, and efforts to explore the latter 
at KEK/J-PARC in Japan. Moreover, there are many other fascinating
aspects of flavour physics, such as charm and top physics, flavour
violation in the charged lepton and neutrino sectors, electric dipole moments 
and studies of the anomalous magnetic moment of the muon. 

Let us in the following discussion focus on the $B$-meson system, which offers
a particularly interesting playground. It involves challenging aspects of
strong interactions, such as issues related to (non)-factorization, rescattering
processes, ``charming penguins", etc., and provides valuable insights into
weak interactions, where the CKM matrix and the associated unitarity triangle
(UT) with its angles $\alpha$, $\beta$ and $\gamma$ are the main targets. 
Moreover, since NP effects may enter this game, there is the exciting possibility 
of obtaining hints for new sources of flavour and/or CP violation.

\subsection{Non-Leptonic $B$ Decays}
A key element for the exploration of CP violation is given by non-leptonic
$B$ decays. The reason is that observable CP-violating effects are induced
by certain interference effects, which may arise in decays of this kind.
The final states of non-leptonic transitions consist only of quarks, and
are caused by $b\to q_1 \bar q_2\,d (s)$ quark-level processes, with 
$q_1,q_2\in\{u,d,c,s\}$. There are two kinds of topologies contributing to 
such decays: ``tree'' and ``penguin'' topologies. The latter consist of gluonic 
(QCD) and electroweak (EW) penguins. Depending on the flavour content of 
their final states, we distinguish between decays which receive {\it only} tree
contributions, channels which may originate from tree {\it and} penguin
contributions, and modes which are {\it only} caused by penguin
topologies. In order to deal with these processes theoretically, low-energy effective
Hamiltonians are used, which are calculated by means of the operator
product expansion, yielding transition amplitudes of the following 
structure \cite{BBL-rev}:
\begin{equation}\label{LEH}
\langle f|{\cal H}_{\rm eff}|B\rangle=
\frac{G_{\rm F}}{\sqrt{2}}\lambda_{\rm CKM}\sum_{k}C_{k}(\mu)\,
\langle f|Q_{k}(\mu)|B\rangle,
\end{equation}
where $G_{\rm F}$ is Fermi's constant, $\lambda_{\rm CKM}$ a factor containing 
the corresponding elements of the CKM matrix, and $\mu$ denotes a 
renormalization scale. The $Q_k$ are local operators, which 
are generated through the interplay between electroweak interactions and QCD, 
and govern ``effectively'' the decay in question, whereas the Wilson coefficients $C_k(\mu)$ describe the scale-dependent ``couplings" of the interaction vertices 
that are associated with the $Q_k$. In this formalism, the short-distance 
contributions are described by the perturbatively calculable Wilson coefficients
$C_{k}(\mu)$, whereas the long-distance physics arises in the form of 
hadronic matrix elements $\langle f|Q_{k}(\mu)|B\rangle$. These non-perturbative
quantities are the key problem in the theoretical analyses of non-leptonic $B$ 
decays. Although there were interesting developments in this field 
through ``QCD factorization" (QCDF) \cite{BBNS}, the ``perturbative 
QCD" (PQCD) approach \cite{PQCD}, ``soft collinear effective theory"
(SCET) \cite{SCET}, and QCD light-cone sum-rule methods \cite{LCSR},
the $B$-factory data indicate that the theoretical challenge remains (see, 
for instance, Refs.~\cite{BFRS}--\cite{CGRS}).

Fortunately, it is possible to circumvent the calculation of the hadronic
matrix elements for the exploration of CP violation:
\begin{itemize}
\item Amplitude relations can be used to eliminate the 
hadronic matrix elements. We distinguish between exact relations, 
using pure ``tree'' decays  of the kind $B\to KD$ or $B_c\to D_sD$, 
and relations, which follow from the flavour symmetries of strong interactions, 
and involve $B_{(s)}\to\pi\pi,\pi K,KK$ modes. 
\item In the neutral $B_q$ systems ($q\in\{d,s\}$), the interference between 
$B^0_q$--$\bar B^0_q$ mixing and decay processes may lead to
``mixing-induced CP violation''. If a single CKM amplitude dominates the 
decay, the hadronic matrix elements cancel in the corresponding CP asymmeties; 
otherwise we have to use amplitude relations again.
\end{itemize}
These avenues offer various strategies to ``overconstrain" the UT through 
studies of CP violation in the $B$-meson system, and to compare the resulting 
picture with the usual ``CKM fits" \cite{CKMfitter,UTfit}. Moreover, ``rare" decays, 
which originate from loop processes in the SM, provide valuable complementary information; key examples are $B\to X_s\gamma$ and the exclusive modes
$B\to K^\ast \gamma$, $B\to\rho\gamma$, as well as $B_{s,d}\to \mu^+\mu^-$ 
and $K^+\to\pi^+\nu\bar\nu$, $K_{\rm L}\to\pi^0\nu\bar\nu$. In the 
presence of NP effects in the quark-flavour sector, we expect discrepancies 
with respect to the pattern following from the structure of the CKM matrix.

\begin{figure}[t]
\centerline{
\begin{tabular}{ll}
  \includegraphics[width=5.7truecm]{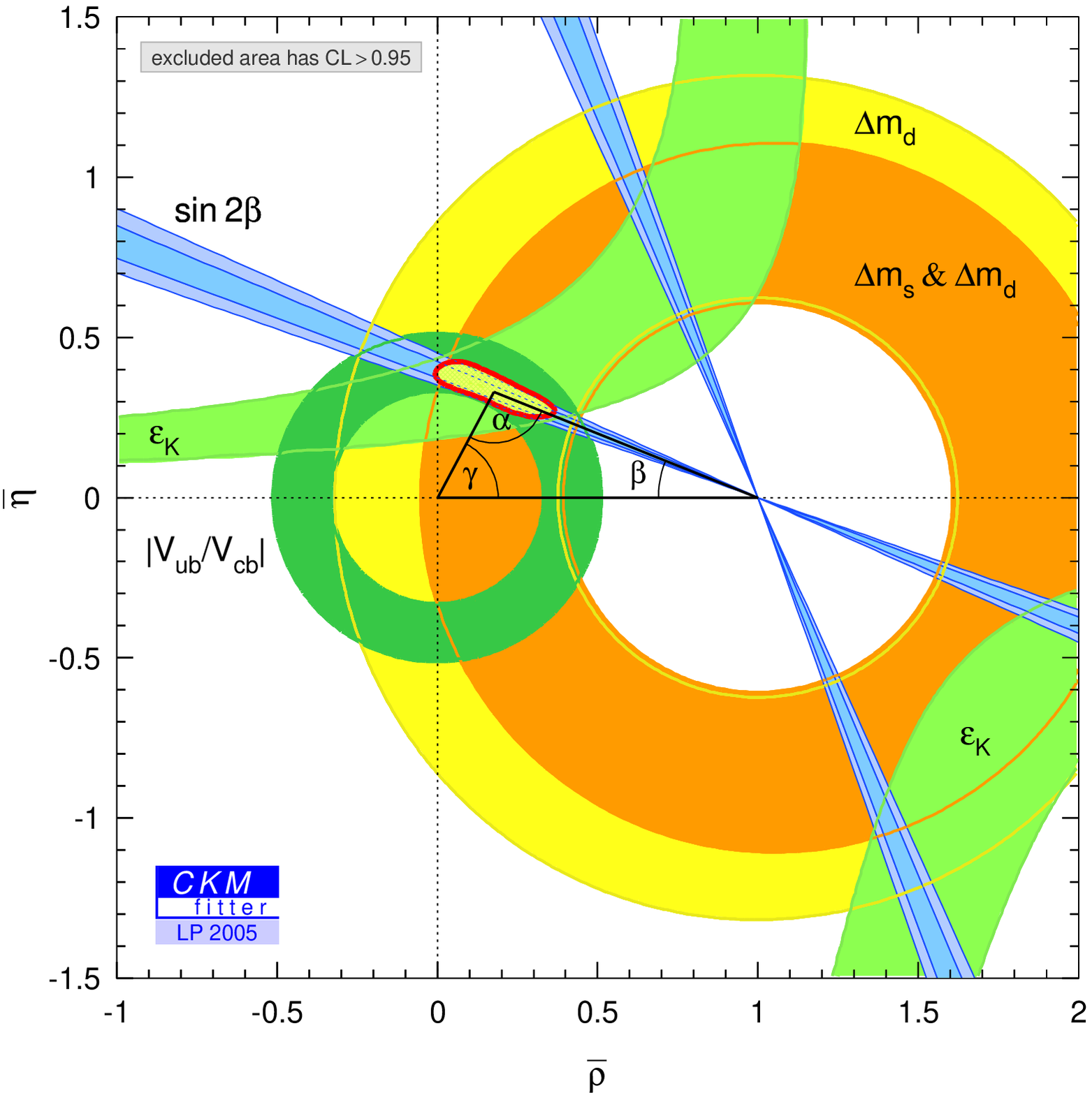} &
\includegraphics[width=6.7truecm]{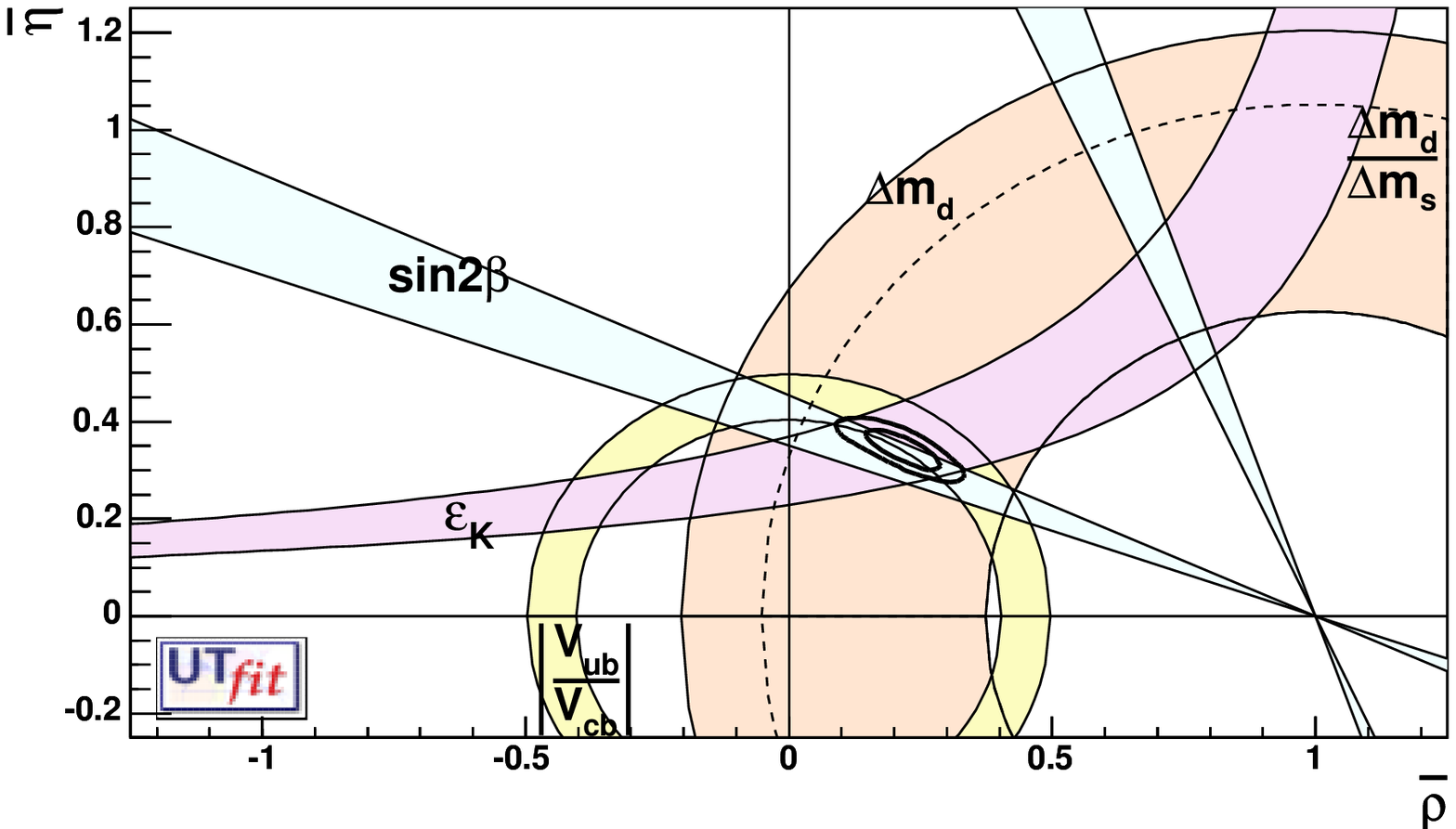}
 \end{tabular}}
\caption{The most recent analyses of the CKMfitter and 
UTfit  collaborations \cite{CKMfitter,UTfit}.}\label{fig:UTfits}
\end{figure}

\subsection{The Generic Impact of New Physics}
Popular avenues for NP to manifest itself are offered by
$B^0_q$--$\bar B^0_q$ mixing and/or decay amplitudes \cite{Botella}.
Let us first have a look at the former option. Here NP may enter through the 
exchange of new particles in box diagrams, which contribute in the SM, or through 
new contributions at the tree level, thereby modifying the mixing parameters as 
follows:
\begin{equation}\label{Dm-Phi-NP}
\Delta M_q=\Delta M_q^{\rm SM}+\Delta M_q^{\rm NP}, \quad
\phi_q=\phi_q^{\rm SM}+\phi_q^{\rm NP}.
\end{equation}
Whereas the NP contribution $\Delta M_q^{\rm NP}$ to the mass difference
of the $B_q$ mass eigenstates would affect the determination of one side,
$R_t$, of the UT, the NP contribution $\phi_q^{\rm NP}$ to the weak mixing
phase would enter the mixing-induced CP asymmetries. The comparison of
the $B$-factory data for the mixing-induced CP violation in the ``golden" decay 
$B^0_d\to J/\psi K_{\rm S}$, which allows a clean measurement of 
$\sin 2\beta$ in the SM, and the CKM fits is globally very good and does not
indicate large NP effects. However, thanks to a new 
Belle result \cite{s2b-belle}, the world average of $(\sin2\beta)_{\psi K_{\rm S}}$ 
compiled by the ``Heavy Flavour Averaging Group" \cite{HFAG} went down by 
about $1 \sigma$ this summer, and takes now the following value:
\begin{equation}\label{s2b-average}
(\sin 2\beta)_{\psi K_{\rm S}}=0.687\pm0.032.
\end{equation} 
Because of this somewhat surprising development, the straight line in the 
$\bar\rho$--$\bar\eta$ plane of the generalized Wolfenstein parameters
\cite{wolf,blo} is now on the lower side of the allowed region for the apex of
the UT following from the CKM fits \cite{CKMfitter,UTfit}, i.e.\ the picture
in the $\bar\rho$--$\bar\eta$ plane does no longer look ``perfect", as can
be seen in Fig.~\ref{fig:UTfits}.
This deviation could be interpreted in terms of NP contributions to 
$B^0_d$--$\bar B^0_d$ mixing, involving in particular a CP-violating phase 
$\phi_d^{\rm NP}\sim -8^\circ$ \cite{UTfit-NP,BFRS-5}. As a next step, it will 
be very interesting to explore the $B_s$ system, where $B^0_s\to J/\psi \phi$ 
offers a sensitive probe to search for CP-violating NP effects in 
$B^0_s$--$\bar B^0_s$ mixing \cite{NiSi}--\cite{DFN}. 

Concerning the possibility for NP to enter directly through decay amplitudes,
the flavour-changing neutral-current (FCNC) sector plays a key r\^ole. For instance, 
new particles may enter in penguin diagrams, or new FCNC processes may arise 
at the tree level. The effects are typically small if the considered decay is
dominated by SM tree processes. This is the case, for example, in 
$B^0_d\to J/\psi K_{\rm S}$. However, this decay receives also contributions from 
penguin topologies, which enter essentially with the same weak phase as the
tree contribution in the SM, and also EW penguins may have a sizeable 
impact \cite{RF-EWP-rev}. Since NP could nicely enter through the latter 
topologies \cite{trojan,FM-BpsiK}, the possible small conflict of (\ref{s2b-average}) 
with the CKM fits could, in principle, also be generated through NP effects in the EW 
penguin sector (or through other NP contributions to the $B\to J/\psi K$ amplitudes), 
although the corresponding contributions would have to be enhanced significantly 
with respect to the SM estimate \cite{RF-landscape}.  A tool to distinguish this 
logical possibility from NP effects in $B^0_d$--$\bar B^0_d$ mixing is offered by 
decays of the kind $B_d\to D\pi^0, D\rho^0, ...$, which are pure ``tree" decays, i.e.\
they do {\it not} receive any penguin contributions. If the neutral $D$ mesons
are observed through their decays into CP eigenstates $D_\pm$, these decays
allow extremely clean determinations of the ``true" value of $\sin2\beta$ 
\cite{RF-BdDpi0}. Consequently, detailed feasibility studies for the exploration 
of the $B_d\to D\pi^0, D\rho^0, ...$\ modes at a super-$B$ factory are strongly 
encouraged. If a decay is dominated by FCNC processes, we may encounter
potentially large NP effects. Interestingly, there are hints in the $B$-factory data 
for such effects. Let us have a closer look at the corresponding puzzles in
the next section.

\section{Prominent Puzzles in the Current $B$-Factory Data}\label{sec:puzzles}
\subsection{CP Violation in $B^0_d\to\phi K_{\rm S}$}\label{ssec:BphiK}
An interesting probe for the testing of the SM description of CP violation is 
offered by the $B_d^0\to \phi K_{\rm S}$ channel, which originates from 
$\bar b\to \bar s s \bar s$ transitions and is, therefore, a 
pure penguin mode. Thanks to the special phase structure of the
corresponding decay amplitude in the SM, the following relations can
be derived \cite{RF-EWP-rev,growo}:
\begin{eqnarray}
{\cal A}_{\rm CP}^{\rm dir}(B_d\to\phi K_{\rm S})&=&0+{\cal O}(\lambda^2)
\label{Adir-rel}\\
\underbrace{{\cal A}_{\rm CP}^{\rm mix}(B_d\to\phi K_{\rm S})}_{\equiv
-(\sin2\beta)_{\phi K_{\rm S}}}&=&
\underbrace{{\cal A}_{\rm CP}^{\rm mix}(B_d\to\psi K_{\rm S})}_{\equiv
-(\sin2\beta)_{\psi K_{\rm S}}}+{\cal O}(\lambda^2),\label{Amix-rel}
\end{eqnarray}
where the ${\cal A}_{\rm CP}^{\rm dir}$ and ${\cal A}_{\rm CP}^{\rm mix}$
denote the ``direct" and ``mixing-induced" CP asymmetries, respectively, 
and $\lambda\equiv|V_{us}|=0.22$ is the Wolfenstein expansion parameter
\cite{wolf}. The dominant contributions to $B\to\phi K$ decays arise from 
QCD penguin operators \cite{BphiK-old}. However, due to the large top-quark 
mass, EW penguins have a sizeable impact as well \cite{RF-EWP,DH-PhiK}. 
Consequently, since $B^0_d\to \phi K_{\rm S}$ is governed by penguin processes 
in the SM, this decay may well be affected by NP. In fact, if we assume that NP 
arises generically in the TeV regime, it can be shown through field-theoretical 
estimates that the NP contributions to  $b\to s\bar s s$ transitions may well lead 
to sizeable violations of the relations in (\ref{Adir-rel}) and (\ref{Amix-rel}) 
\cite{FM-BphiK}. Moreover, this is also the case for several specific NP scenarios
(see, for instance, Refs.~\cite{CFMS}--\cite{GHK}). 

Concerning the measurement of the CP asymmetries of the $B^0_d\to\phi K_{\rm S}$
decay, the result $(\sin2\beta)_{\phi K_{\rm S}}=-0.96\pm0.50^{+0.09}_{-0.11}$ 
reported by the Belle collaboration in the summer of 2003 led to quite some 
excitement in the $B$-physics community. Meanwhile, the Babar
\cite{BaBar-Bphi-K} and Belle \cite{Belle-Bphi-K} results are in good agreement 
with each other, yielding the following averages \cite{HFAG}:
\begin{equation}\label{BphiK-av}
{\cal A}_{\rm CP}^{\rm dir}(B_d\to \phi K_{\rm S})=-0.09\pm0.14, \quad
(\sin2\beta)_{\phi K_{\rm S}}=0.47\pm0.19. 
\end{equation}
The number for $(\sin2\beta)_{\phi K_{\rm S}}$
is still on the lower side, and may indicate NP contributions
to $b\to s\bar ss$ processes. In \cite{RF-landscape}, a much more detailed
discussion can be found, addressing also NP effects in the EW penguin sector, 
which may be responsible both for the possible difference between 
(\ref{s2b-average}) and (\ref{BphiK-av}) and for the ``$B\to\pi K$ puzzle"
discussed in Subsection~\ref{ssec:BpiK}.

It will be very interesting to follow the evolution of the $B$-factory data,
and to monitor also similar modes, such as $B^0_d\to \pi^0 K_{\rm S}$ 
\cite{PAPIII} and  $B_d^0\to \eta'K_{\rm S}$ \cite{loso}. For a compilation of 
the corresponding newest experimental results, see Ref.~\cite{HFAG}; 
recent theoretical papers dealing with these channels can be found in 
Refs.~\cite{BFRS,BFRS-5,GGR,beneke,SCET-Bdpi0K0}.

\subsection{The $B\to\pi K$ Puzzle and its Relation to Rare Decays}\label{ssec:BpiK}
\subsubsection{Preliminaries}
The first indication of the ``$B\to\pi K$ puzzle" goes back to the observation of 
the $B^0_d\to\pi^0K^0$ channel by the CLEO collaboration in 2000 with a 
remarkably prominent rate, which may signal a discrepancy 
with the SM \cite{BF00}. This possible anomaly is still present in the 
current data, and has recently received a lot of attention (see, for instance,
\cite{BCLL-puzzle,BeNe,yoshikawa,GR03,MY,WZ}). 

Let us follow here the strategy to explore this exciting topic in a systematic manner 
that was developed in \cite{BFRS}. The starting point is an analysis of the 
$B\to\pi\pi$ system, where the data can be accommodated in the SM through 
large non-factorizable effects. In particular, the $B\to\pi\pi$ decays allow the 
extraction of a set of hadronic parameters  with the help of the isospin symmetry 
of strong interactions. Using then the $SU(3)$ flavour symmetry and neglecting 
certain exchange and penguin annihilation topologies, the hadronic $B\to\pi\pi$
parameters can be converted into their $B\to\pi K$ counterparts, allowing the 
prediction of all $B\to\pi K$ observables in the SM. Interestingly, agreement 
with experiment is found for those decays  that are only marginally affected 
by (colour-suppressed) EW penguins. On the other hand, the SM predictions 
of the $B\to\pi K$ observables which are significantly affected by (colour-allowed) 
EW penguins are not found in agreement with the data, thereby reflecting the 
$B\to\pi K$ puzzle. Moreover, internal consistency checks of the working assumptions 
can be performed, which work well within the current uncertainties, and the 
numerical results turn out to be very stable with respect to large 
non-factorizable $SU(3)$-breaking corrections \cite{BFRS-5}. 
In view of these features, NP in the EW penguin sector may be at the origin of the 
$B\to\pi K$ puzzle. In fact, it can be resolved through 
a modification of the EW penguin parameters, involving in particular a large 
CP-violating NP phase that vanishes in the SM. The implications
of this kind of NP on rare $K$ and $B$ decays are then investigated in the final
step of the strategy proposed in Ref.~\cite{BFRS}.

The numerical results presented below refer to the very recent analysis
presented in Ref.~\cite{BFRS-5}. In view of the new world average for
$(\sin 2\beta)_{\psi K_{\rm {S}}}$ in (\ref{s2b-average}), which may signal
NP effects in $B^0_d$--$\bar B^0_d$ mixing, the CP asymmetries  of the 
$B^0_d\to\pi^+\pi^-$, $B^0_d\to\pi^-K^+$ system are used to determine the 
``true" value of the UT angle $\gamma$, yielding
\begin{equation}\label{gamma-det}
\gamma=(73.9^{+5.8}_{-6.5})^\circ.
\end{equation}
This number is then used as an input for the $B\to\pi\pi,\pi K$ analysis discussed 
below. Furthermore, complementing (\ref{gamma-det}) with the experimental value 
of the UT side $R_b\propto|V_{ub}/V_{cb}|$, which follows from semi-leptonic $B$ 
decays that are very robust under NP effects, also the ``true" value of $\beta$ 
can be extracted, $\beta=(25.8\pm 1.3)^\circ$, which would correspond to 
$\phi_d^{\rm NP}=-(8.2\pm 3.5)^\circ$,  in accordance 
with the analysis of Ref.~\cite{UTfit-NP}.

\subsubsection{The $B\to\pi\pi$ Analysis}
The starting point of the $B\to\pi\pi$ study is given by the following ratios:
\begin{eqnarray}
R_{+-}^{\pi\pi}&\equiv&2\left[\frac{\mbox{BR}(B^\pm\to\pi^\pm\pi^0)}{\mbox{BR}
(B_d\to\pi^+\pi^-)}\right]=F_1(d,\theta,x,\Delta;\gamma)
\stackrel{\rm exp}{=}2.04\pm0.28
\label{Rpm-def}\\
R_{00}^{\pi\pi}&\equiv&2\left[\frac{\mbox{BR}(B_d\to\pi^0\pi^0)}{\mbox{BR}
(B_d\to\pi^+\pi^-)}\right]=F_2(d,\theta,x,\Delta;\gamma)
\stackrel{\rm exp}{=}0.58\pm0.13.
\end{eqnarray}
Here the isospin symmetry of strong interactions was used to express
these observables in terms of $\gamma$ and the hadronic parameters $de^{i\theta}$,
$xe^{i\Delta}$ which were introduced in Ref.~\cite{BFRS}. Moreover, the 
CP asymmetries
\begin{eqnarray}
{\cal A}_{\rm CP}^{\rm dir}(B_d\to \pi^+\pi^-)&=&
G_1(d,\theta;\gamma)\stackrel{\rm exp}{=}-0.37\pm0.10 \label{Adirpi+pi-}\\
{\cal A}_{\rm CP}^{\rm mix}(B_d\to \pi^+\pi^-)&=&
G_2(d,\theta;\gamma,\phi_d)\stackrel{\rm exp}{=}+0.50\pm0.12\label{Amixpi+pi-}
\end{eqnarray}
are at our disposal, where $\phi_d\stackrel{\rm exp}{=}(43.4\pm2.5)^\circ$
follows from the data for the mixing-induced CP violation in 
$B^0_d\to J/\psi K_{\rm S}$. Using the value of $\gamma$ in (\ref{gamma-det}), 
the hadronic parameters characterizing the $B\to\pi\pi$ system can be extracted, 
with the following results:
\begin{equation}\label{d-theta-x-Delta}
d=0.52^{+0.09}_{-0.09}, \quad 
\theta=(146^{+7.0}_{-7.2})^\circ, \qquad
x=0.96^{+0.13}_{-0.14}, \quad 
\Delta=-(53^{+18}_{-26})^\circ.
\end{equation}
These numbers, which exhibit large non-factorizable effects, take also 
EW penguin effects into account, although these topologies 
have a minor impact on the $B\to\pi\pi$ decays. 

Finally, the CP-violating observables of the $B^0_d\to\pi^0\pi^0$ channel 
can be predicted: 
\begin{eqnarray}
{\cal A}_{\rm CP}^{\rm dir}(B_d\to \pi^0\pi^0)&=&-0.30^{+0.48}_{-0.26}
\,\stackrel{\rm exp}{=}\, -0.28^{+0.40}_{-0.39}\label{Adir00-pred}\\
{\cal A}_{\rm CP}^{\rm mix}(B_d\to \pi^0\pi^0)&=&-0.87^{+0.29}_{-0.19}.
\end{eqnarray}
Although no stringent test of these predictions is possible at this stage, the 
indicated agreement between the prediction in (\ref{Adir00-pred}) and the
corresponding experimental value \cite{HFAG} is encouraging.

\subsubsection{The $B\to\pi K$ Analysis}
Using now the $SU(3)$ flavour symmetry and neglecting exchange and 
penguin annihilation topologies, the hadronic parameters in (\ref{d-theta-x-Delta})
can be converted into their $B\to\pi K$ counterparts, allowing the prediction of
the $B\to\pi K$ observables in the SM. Moreover, a couple of internal
consistency checks of these working assumptions can be performed, which
are fulfilled by the current data, and the sensitivity of the SM predictions
on large non-factorizable $SU(3)$-breaking effects turns out to be surprisingly
small \cite{BFRS-5}. Consequently, no anomaly is indicated in this sector. 

\begin{figure}
\begin{center}
\includegraphics[width=8.3cm]{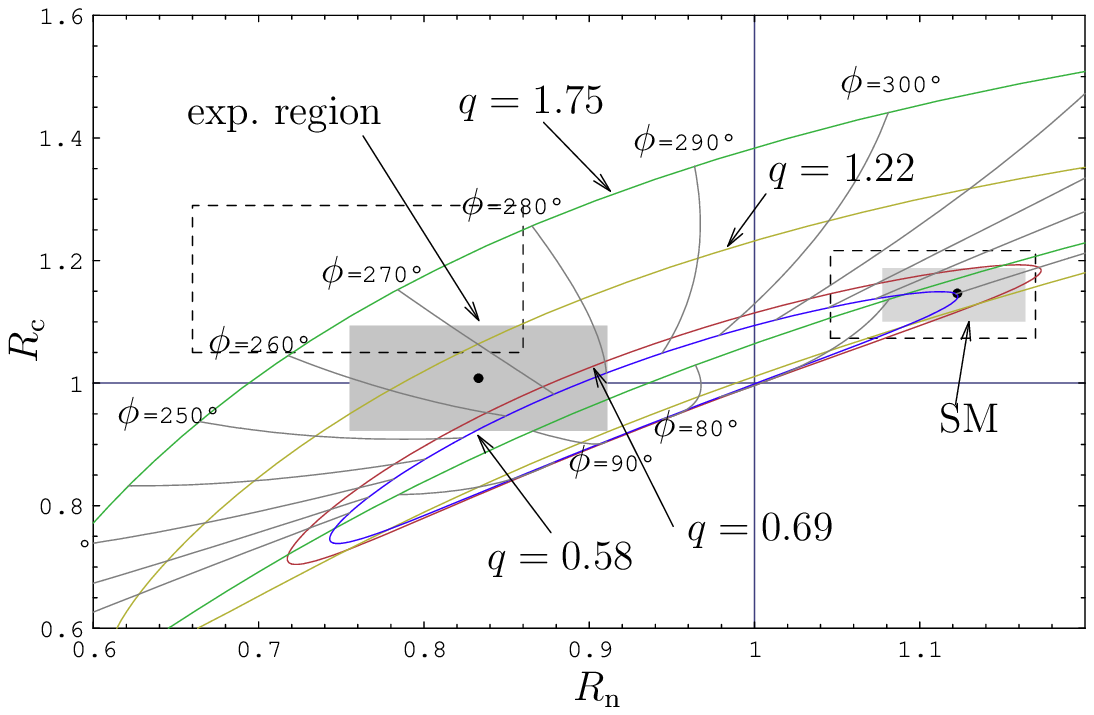}
\end{center}
\vspace*{-0.3truecm}
\caption{The situation in the $R_{\rm n}$--$R_{\rm c}$ plane, as discussed
in the text.}\label{fig:RnRc}
\end{figure}

In the case of the $B^0_d\to\pi^-K^+$, $B^+\to\pi^+K^0$ system, where 
EW penguins have a minor impact, a picture arises in the SM that is in 
accordance with the data (see also \cite{BFRS-up}). 
In order to analyze the decays $B^+\to\pi^0K^+$ and 
$B^0_d\to\pi^0K^0$, which are significantly affected by EW penguins,
it is useful to introduce the following quantities:
\begin{eqnarray}
R_{\rm c}&\equiv&2\left[\frac{\mbox{BR}(B^\pm\to\pi^0K^\pm)}{\mbox{BR}
(B^\pm\to\pi^\pm K)}\right] \,\stackrel{\rm exp}{=}\, 1.01\pm 0.09\\
R_{\rm n}&\equiv&\frac{1}{2}\left[\frac{\mbox{BR}(B_d\to\pi^\mp K^\pm)}{\mbox{BR}
(B_d\to\pi^0K)}\right] \,\stackrel{\rm exp}{=}\, 0.83\pm0.08.
\end{eqnarray}
The EW penguin effects are described by a parameter $q$, which measures
the strength of the EW penguins with respect to the tree-diagram-like topologies,
and a CP-violating phase $\phi$. In the SM, this phase vanishes,
and $q$ can be calculated with the help of the $SU(3)$ flavour symmetry,
yielding a value of $0.69\times 0.086/|V_{ub}/V_{cb}|=0.58$ \cite{NR}. The 
situation can transparently be discussed in the 
$R_{\rm n}$--$R_{\rm c}$ plane, as shown in Fig.~\ref{fig:RnRc}:
the shaded areas indicate the SM prediction and the experimental range, 
the lines show the theory predictions for the central values of the hadronic 
parameters and various values of $q$ with $\phi\in[0^\circ,360^\circ]$;
the dashed rectangles represent the SM predictions and experimental ranges 
at the time of the original analysis of Ref.~\cite{BFRS}.  Although the central values of 
$R_{\rm n}$ and $R_{\rm c}$ have slightly moved towards each other, the 
puzzle is as prominent as ever. 
The experimental region can now be reached without an enhancement of $q$, 
but a large CP-violating phase $\phi$ of the order of $-90^\circ$ is
still required, although $\phi\sim+90^\circ$ can also bring us rather close
to the experimental range of $R_{\rm n}$ and $R_{\rm c}$.

Moreover, also the CP-violating asymmetries of the 
$B^\pm\to\pi^0 K^\pm$ and $B_d\to\pi^0K_{\rm S}$ decays can be predicted
both in the SM and in the scenario of NP effects in the EW penguin sector. 
In particular the mixing-induced CP asymmetry of the latter decay has recently 
received a lot of attention, as the current $B$-factory data give a value of 
\begin{equation}
\Delta S \equiv (\sin2\beta)_{\pi^0K_{\rm S}}-(\sin2\beta)_{\psi K_{\rm S}}
\stackrel{{\rm exp}}{=}-0.38\pm 0.26. 
\end{equation}
In the strategy described above, this difference is predicted to be {\it positive}
in the SM, and in the ballpark of $0.10$--$0.15$ \cite{BFRS-5}. Interestingly,
the best values for $(q,\phi)$ that are implied by the measurements of $R_{\rm n,c}$ 
make the disagreement of $\Delta S$ with the data even larger than in the SM. 
However, also values of $(q,\phi)$ can be found for which $\Delta S$ could be 
smaller than in the SM or even reverse the sign \cite{BFRS-5}. This happens in
particular for $\phi\sim+90^\circ$, i.e.\ if the CP-violating NP phase flips its sign. 
In this case, also the central value of $(\sin2\beta)_{\phi K_{\rm S}}$ in (\ref{BphiK-av})
could be straightforwardly accommodated in this scenario of NP \cite{RF-landscape},
and could in fact be another manifestation of a modified EW penguin sector
with new sources for CP violation.

\subsubsection{Relation with Rare $B$ and $K$ Decays}
A popular scenario for NP effects in the EW penguin sector is
offered by modified $Z^0$ penguins with a new CP-violating phase. 
This scenario was already considered in the literature, where 
model-independent analyses and studies within SUSY can be found 
\cite{Z-pen-analyses,BuHi}. Following \cite{BFRS-I} and performing a
renormalization-group evolution from scales ${\cal O}(m_b)$ to
${\cal O}(M_W,m_t)$, the EW penguin parameters $(q,\phi)$ of the 
$B\to\pi K$ system can be converted, in this scenario, into a $Z^0$-penguin 
function $C$ as well as other short-distance functions, which allow us 
to make predictions for rare decays the kind $K^+\to\pi^+\nu\bar\nu$, 
$K_{\rm L}\to\pi^0\nu\bar\nu$, $K_{\rm L}\to\pi^0\ell^+\ell^-$, 
$B\to X_s\nu \bar\nu$ and $B_{s,d}\to\mu^+\mu^-$. An analysis along
these lines shows that we may encounter interesting NP effects in
the corresponding observables, in particular
in the $K\to\pi\nu\bar\nu$ system. In \cite{BFRS-5}, it was pointed out
that the most recent $B$-factory constraints for rare decays, in particular
for $B\to X_s\ell^+\ell^-$ \cite{Bobeth:2005ck}, have interesting new implications.
In this context, a few future scenarios with different patterns of
the relevant observables are discussed, where also the mixing-induced
CP violation in $B^0_d\to\pi^0K_{\rm S}$ plays a prominent r\^ole. It will
be interesting to confront this analysis with future, more accurate data!

\begin{figure}
\vspace*{0.3truecm}
\begin{center}
\includegraphics[width=9.5cm]{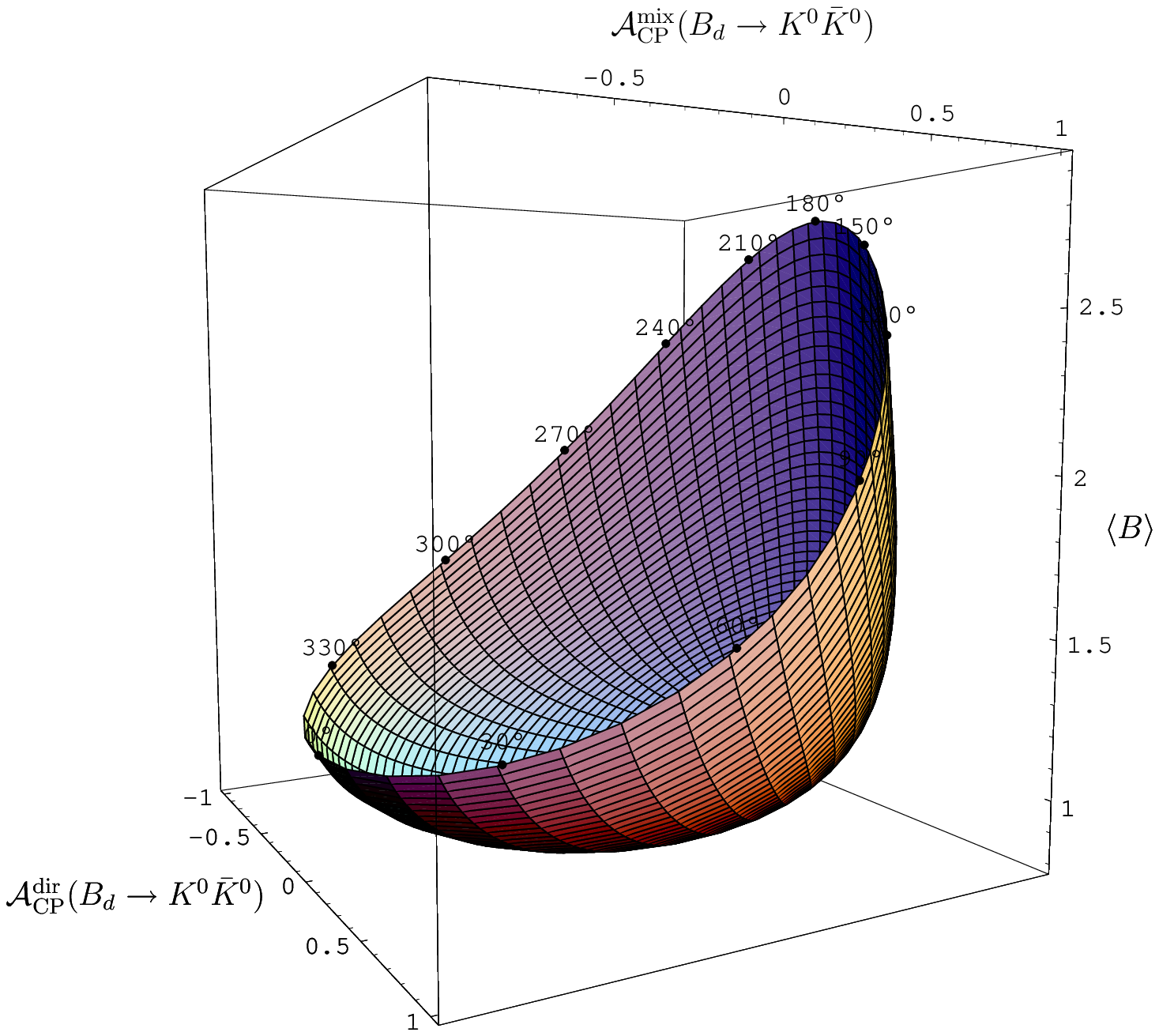}
\end{center}
\vspace*{-0.3truecm}
\caption{Illustration of the surface in the 
${\cal A}_{\rm CP}^{\rm dir}$--${\cal A}_{\rm CP}^{\rm mix}$--$\langle B \rangle$
observable space characterizing the $B^0_d\to K^0\bar K^0$ decay in the SM. 
The intersecting lines on the surface correspond to constant 
values of $\rho_{K\!K}$ and $\theta_{K\!K}$; the numbers on the fringe indicate 
the value of $\theta_{K\!K}$, while the fringe itself is defined by 
$\rho_{K\!K}=1$.}\label{fig:SM-surface}
\end{figure}

\section{Entering New Territory: $b\to d$ Penguins}\label{sec:b-d-pen}
Another recent hot topic is the exploration of $b\to d$ penguin processes. 
Both the non-leptonic decays belonging to this category, which originate
from $b\to d \bar s s$ quark transitions, and the radiative decays
caused by  $b\to d\gamma$ processes are now coming within experimental 
reach at the $B$ factories. We are therefore entering a new territory, which 
is still essentially unexplored.

\subsection{A Prominent Example: $B^0_d\to K^0\bar K^0$}
This decay is the $\bar b\to \bar d$ penguin counterpart of the $B^0_d\to\phi K_{\rm S}$
decay discussed in Subsection~\ref{ssec:BphiK}. The dominant r\^ole is played 
by QCD penguins; since EW penguins may only contribute in colour-suppressed 
form, they have a minor impact on $B^0_d\to K^0\bar K^0$, in contrast to the case 
of $B^0_d\to\phi K^0$, where they may also contribute in colour-allowed form. 
In the SM, the $B^0_d\to K^0\bar K^0$ decay amplitude can be written as follows:
\begin{equation}\label{ampl-BdKK-lamt}
A(B^0_d\to K^0\bar K^0)=\lambda^3A(A_{\rm P}^t-A_{\rm P}^c)
\left[1-\rho_{K\!K} e^{i\theta_{K\!K}}e^{i\gamma}\right],
\end{equation}
where $\lambda$ and $A$ are Wolfenstein parameters \cite{wolf}, and the
hadronic parameter
\begin{equation}\label{rho-KK-def}
\rho_{K\!K} e^{i\theta_{K\!K}}\equiv R_b
\left[\frac{A_{\rm P}^t-A_{\rm P}^u}{A_{\rm P}^t-A_{\rm P}^c}\right]
\end{equation}
involves the side $R_b$ of the UT and the strong amplitudes 
$A_{\rm P}^q$, which describe penguin topologies with $q$-quark exchanges.
The direct and mixing-induced CP asymmetries of $B^0_d\to K^0\bar K^0$
can then be expressed in terms of $\rho_{K\!K}$, $\theta_{K\!K}$ and $\gamma$;
the latter observable involves also the mixing phase $\phi_d$. If we assume, for
a moment, that the penguin contributions are dominated 
by top-quark exchanges, (\ref{rho-KK-def}) simplifies as 
$\rho_{K\!K} e^{i\theta_{K\!K}} \to R_b$. Since the CP-conserving strong phase 
$\theta_{K\!K}$ vanishes in this limit,
the direct CP violation in $B^0_d\to K^0\bar K^0$ vanishes, too. Moreover, 
it can be shown that also the mixing-induced CP asymmetry would vanish
in the SM because of $\phi_d=2\beta$. Consequently, the measurement of
the CP-violating $B^0_d\to K^0\bar K^0$ asymmetries appears as an interesting
test of the SM (see, for instance, \cite{quinn}). However, contributions from 
penguins with up- and charm-quark exchanges are expected to yield 
sizeable CP violation in $B_d^0\to K^0\bar K^0$ even in the SM, so that
the interpretation of these effects is much more complicated \cite{RF-BdKK}.
Moreover, these contributions contain also possible long-distance rescattering 
effects \cite{BFM}, so that  $\rho_{K\!K}$ and $\theta_{K\!K}$ are affected
by large uncertainties.

Despite this problem, interesting insights can be obtained through the
$B^0_d\to K^0\bar K^0$ observables \cite{FR1}. If we keep 
$\rho_{K\!K}$ and $\theta_{K\!K}$ as free parameters, we may
characterize this decay in the SM through a surface in
observable space, which is shown in Fig.~\ref{fig:SM-surface}. 
It should be emphasized that this surface is {\it theoretically clean} since it 
relies only on the general SM parametrization of $B^0_d\to K^0\bar K^0$. 
Consequently, should future measurements give a value in observable space 
that should {\it not} lie on the SM surface, we would have immediate evidence 
for NP contributions to $\bar b\to \bar d s \bar s$ FCNC processes. 
However, while the direct and mixing-induced CP asymmetries can be 
``straightforwardly" determined through time-dependent rate asymmetries, the 
extraction of $\langle B\rangle$ from the CP-averaged branching ratio requires 
further input:
\begin{equation}\label{BR-BKK-expr}
\mbox{BR}(B_d\to K^0\bar K^0)=\frac{\tau_{B_d}}{16\pi M_{B_d}}
\times \Phi_{KK} \times
|\lambda^3 A \, A_{\rm P}^{tc}|^2 \langle B \rangle,
\end{equation}
where $\Phi_{KK}$ denotes a two-body phase-space factor and
$A_{\rm P}^{tc}\equiv A_{\rm P}^t-A_{\rm P}^c$.
In order to fix the overall normalization factor involving the penguin 
amplitude $ A_{\rm P}^{tc}$,  we may either use (i) $B\to\pi\pi$ ($b\to d$), 
or (ii) $B\to\pi K$ ($b\to s$) decays.

As can be seen in Fig.~\ref{fig:SM-surface}, $\langle B\rangle$ takes an absolute
minimum, 
\begin{equation}\label{B-bound}
\langle B \rangle\equiv 1-2\rho_{K\!K}\cos\theta_{K\!K}
\cos\gamma+\rho_{K\!K}^2 \, \geq \, \sin^2\gamma,
\end{equation}
which can be converted into the following lower bounds for the CP-averaged
$B_d\to K^0 \bar K^0$ branching ratio:
\begin{equation}\label{BdKK-bounds}
\mbox{BR}(B_d\to K^0\bar K^0)
\geq
\left\{\begin{array}{ll}
(1.39^{+1.54}_{-0.95})\times 10^{-6} & \mbox{(i),}\\
(1.36^{+0.18}_{-0.21})\times 10^{-6} & \mbox{(ii).}
\end{array}\right.
\end{equation}
Interestingly, both avenues to fix the overall normalization through $SU(3)$ 
flavour-symmetry arguments give results in nice agreement with each other. At the 
time of the derivation of these bounds, the $B$ factories reported an 
experimental {\it upper} bound of $\mbox{BR}(B_d\to K^0\bar K^0)<1.5\times 10^{-6}$ (90\% C.L.). Consequently, the theoretical {\it lower} bounds given above
suggested that the observation of this channel should just be ahead of us. 
Subsequently, the first signals were indeed announced, in accordance with 
(\ref{BdKK-bounds}):
\begin{equation}\label{BdK0K0-data}
\mbox{BR}(B_d\to K^0\bar K^0)=\left\{
\begin{array}{ll}
(1.19^{+0.40}_{-0.35}\pm0.13) \times 10^{-6} & \mbox{(BaBar 
\cite{BaBar-BKK}),}\\
 (0.8\pm0.3\pm0.1) \times 10^{-6} & \mbox{(Belle \cite{Belle-BKK}).}
 \end{array}\right.
\end{equation}
The SM description of $B^0_d\to K^0\bar K^0$ has thus successfully passed its 
first test. However, the experimental errors are still very large, and the next crucial 
step -- a measurement of the CP asymmetries -- is still missing. For further aspects 
of $B^0_d\to K^0\bar K^0$ and a discussion of $SU(3)$-breaking effects, 
the reader is referred to Ref.~\cite{FR1}.

\subsection{General Lower Bounds for $b\to d$ Penguin Processes}
The bounds discussed above are actually realizations of a general, 
model-independent bound that can be derived in the SM for $b\to d$ 
penguin processes \cite{FR2}. If we consider such a decay, $\bar B \to \bar f_d$, 
we may -- in analogy to (\ref{ampl-BdKK-lamt}) -- write its amplitude as follows:
\begin{equation}
A(\bar B \to \bar f_d)= A^{(0)}_d
\left[1-\rho_de^{i\theta_d}e^{-i\gamma}\right].
\end{equation}
Keeping $\rho_d$ and $\theta_d$ as ``unknown", i.e.\ free hadronic parameters
yields
\begin{equation}
\mbox{BR}(B \to f_d)\geq\tau_B
\left[\sum_{\rm Pol}\int \!\! d \, {\rm PS} \, |A^{(0)}_d|^2 \right]\sin^2\gamma,
\end{equation}
where we made the phase-space integration explicit and the 
sum runs over possible polarization configurations of the final state $f_d$. 
In order to deal with the term in square brackets, we use a $b\to s$ 
penguin decay $\bar B \to \bar f_s$, which is the counterpart of $\bar B \to \bar f_d$ 
in that the corresponding CP-conserving strong amplitudes can be related
to one another through the $SU(3)$ flavour symmetry. We may then write
\begin{equation}
A(\bar B \to \bar f_s)= - \frac{A^{(0)}_s}{\sqrt{\epsilon}}
\left[1+\epsilon\rho_s e^{i\theta_s}e^{-i\gamma}\right],
\end{equation}
where $\epsilon\equiv\lambda^2/(1-\lambda^2)=0.05$. Neglecting the
term proportional to $\epsilon$ in the square bracket, which gives only a 
small correction at the percent level, we arrive at
\begin{equation}\label{general-bound}
\frac{\mbox{BR}(B \to f_d)}{\mbox{BR}(B \to f_s)}
\geq \epsilon \left[\frac{\sum_{\rm Pol}\int \! d \, {\rm PS} \, 
|A^{(0)}_d|^2 }{\sum_{\rm Pol}\int \! d \, {\rm PS} \, |A^{(0)}_s|^2 }
\right]\sin^2\gamma
\stackrel{SU(3)_{\rm F}}{\longrightarrow} \epsilon \sin^2\gamma.
\end{equation}
Since $\sin^2\gamma$ is favourably large in the SM and the decay
$\bar B \to \bar f_s$ will be measured before its $b\to d$ 
counterpart  -- simply because of the CKM enhancement -- 
(\ref{general-bound}) provides strong lower bounds for 
$\mbox{BR}(B \to f_d)$. 

Let us now discuss applications of (\ref{general-bound}), where
$SU(3)$-breaking effects are included in the numerical results, 
as discussed in detail in \cite{FR2}. Concerning non-leptonic
decays, the following picture emerges:
\begin{equation}\label{BKK-numbers}
\begin{array}{llll}
1.69\,^{+0.21}_{-0.24} & \leq & \mbox{BR}(B^\pm\!\to\! K^\pm K)/10^{-6} &
\stackrel{{\rm exp}}{<} 2.4,  \\
0.68\,^{+0.11}_{-0.13} & \leq &\mbox{BR}(B^\pm\!\to\! K^\pm K^\ast)/10^{-6} &
\stackrel{{\rm exp}}{<} 5.3, \\
0.64^{+0.15}_{-0.16} & \leq & \mbox{BR}(B^\pm\to K^{\ast\pm} K^\ast)/10^{-6} & 
\stackrel{{\rm exp}}{<} 71 .
\end{array}
\end{equation}
This summer, the following new results were reported:
\begin{equation}
\mbox{BR}(B^\pm\!\to\! K^\pm K)=
\left\{\begin{array}{ll}
(1.5\pm0.5\pm0.1)\times 10^{-6} & \mbox{(BaBar \cite{BaBar-BKK})}\\
(1.0\pm0.4\pm0.1)\times 10^{-6} & \mbox{(Belle \cite{Belle-BKK}),}
\end{array} \right.
\end{equation}
which complement (\ref{BdK0K0-data}), and are the first evidence for the 
$B^\pm\!\to\! K^\pm K$ channel, in accordance with the corresponding 
bound in (\ref{BKK-numbers}). In the case of the other modes, the experimental
upper bounds still leave a lot of space. Searches of these decays at the $B$ factories
are strongly encouraged. 

Another important tool to explore the $b\to d$ penguin sector is provided by 
$\bar B\to\rho\gamma$ modes. In this case, the following
picture emerged \cite{FR2}:
\begin{equation}\label{Brhogam-numbers}
\begin{array}{llll}
1.02\,^{+0.27}_{-0.23} & \leq &
\mbox{BR}(B^\pm\to \rho^\pm\gamma)/10^{-6} & 
\stackrel{{\rm exp}}{<} 1.8, \\
0.51\,^{+0.13}_{-0.11} & \leq &
\mbox{BR}(B_d\to \rho^0\gamma)/10^{-6} & 
\stackrel{{\rm exp}}{<} 0.4.
\end{array}
\end{equation}
Consequently, it was expected that $\bar B\to\rho\gamma$ modes should 
soon be discovered at the $B$ factories. Indeed, the Belle 
collaboration reported recently the observation of $b\to d\gamma$ processes, 
with the following results \cite{Belle-bdgam-obs}:
\begin{eqnarray}
\mbox{BR}(B^\pm\to \rho^\pm\gamma)&=&\left(0.55^{+0.43+0.12}_{-0.37-0.11}\right)
\times 10^{-6},
\label{Belle-Brhogam-p}\\
\mbox{BR}(B_d\to \rho^0\gamma)&=&\left(1.17^{+0.35+0.09}_{-0.31-0.08}\right)
\times10^{-6},
\label{Belle-Brhogam-n}\\
\mbox{BR}(B\to(\rho,\omega)\gamma)&=&
\left(1.34^{+0.34+0.14}_{-0.31-0.10}\right)\times10^{-6},
\end{eqnarray}
which was one of the hot topics of the 2005 summer conferences \cite{Belle-press}.
These measurements suffer still from large uncertainties, and the pattern of the 
central values of (\ref{Belle-Brhogam-p}) and (\ref{Belle-Brhogam-n}) would be in
contrast to the expectation following from the isospin symmetry. It will be interesting
to follow the evolution of the data. The next important conceptual step would be the measurement of the corresponding CP-violating observables, though this is still
in the distant future. In view of 
\begin{equation}\label{Bpi-ellell-bounds}
\mbox{BR}(B^\pm\to \pi^\pm \ell^+\ell^-), \,
\mbox{BR}(B^\pm\to \rho^\pm \ell^+\ell^-) \,
\mathrel{\hbox{\rlap{\hbox{\lower4pt\hbox{$\sim$}}}\hbox{$>$}}}
\, 10^{-8},
\end{equation}
a similar comment applies to this species of $b\to d$ penguin decays.

\subsection{Comments}
The stringent experimental upper bounds and the emerging signals for the
$B^\pm\to K^\pm K$ decays disfavour large rescattering effects, which has
important implications for the analysis of the $B\to\pi K$ system \cite{BFRS-up}.

Concering the radiative $B\to \rho\gamma$ modes, an interesting
alternative to confront the data for the branching ratios with the SM is 
offered by converting them into information on the UT side $R_t\propto |V_{td}|$. 
Such an analysis was recently performed by the authors of Refs.~\cite{ALP-rare,BoBu}, 
who also used the $SU(3)$ flavour symmetry, but calculated the CP-conserving
(complex) parameter $\delta a$ entering 
$\rho_{\rho\gamma}e^{i\theta_{\rho\gamma}}=R_b\left[1+\delta a\right]$
in the QCDF approach. The corresponding result, which favours a small impact 
of $\delta a$, takes leading and next-to-leading order QCD corrections into 
account and holds to leading order in the heavy-quark limit \cite{BoBu}. However,
in view of the remarks about possible long-distance effects made above and the 
$B$-factory data for the $B\to\pi\pi$ system, which indicate large corrections 
to the QCDF picture for non-leptonic $B$ decays into two light pseudoscalar 
mesons, it is not obvious that the impact of $\delta a$ is actually small. The 
advantage of the bounds given above is that they are  -- by construction -- {\it not} 
affected by $\rho_{\rho\gamma}e^{i\theta_{\rho\gamma}}$ at all.

Instead of confirming the bounds, it would of course be much more exciting
if some of them were significantly violated through the impact of NP contributions,
interfering destructively with the SM amplitudes. The $b \to d$ penguin decay 
classes are governed by different operators:
\begin{itemize}
\item Non-leptonic decays: four-quark operators.
\item $b\to d \gamma$:
\,$Q_{7,8}^{d}=\frac{1}{8\pi^2}m_b\bar d_i \sigma^{\mu\nu}(1+\gamma_5)
\left\{e b_i F_{\mu\nu} ,\, g_{\rm s}T^a_{ij}b_j G^a_{\mu\nu} \right\}$.
\item $b\to d \ell^+\ell^-$:
\,$Q_{9,10}=\frac{\alpha}{2\pi}(\bar\ell\ell)_{\rm V\!,\,A}
(\bar d_i b_i)_{\rm V-A}$.
\end{itemize}
Consequently, we may actually encounter surprises.

\section{The ``El Dorado" for the LHC: $B_s$ Decays}\label{sec:Bs}
\subsection{Basic Features}
Another essentially unexplored territory is given by the $B_s$-meson system,
since no $B_s$ mesons are accessible at the $e^+e^-$ $B$ factories operating
at the $\Upsilon(4S)$ resonance. However, plenty of $B_s$ mesons are
produced at hadron colliders, i.e.\ at the Tevatron and later on at the LHC 
\cite{Nakada}. Already the $B^0_s$--$\bar B^0_s$ mixing parameters are
particularly interesting quantities. The mass difference $\Delta M_s$
can be combined with its $B_d$-meson counterpart $\Delta M_d$, allowing
the determination of the UT side $R_t$ with the help of a hadronic parameter
$\xi$, which equals one in the strict $SU(3)$ flavour-symmetry limit. The uncetainties
of $\xi$ are an important aspect of lattice QCD studies \cite{lattice}. So far,
$B^0_s$--$\bar B^0_s$ oscillations could not yet be observed, and only 
lower bounds for $\Delta M_s$ are available from the data of the LEP experiments,
SLD and the Tevatron. The most recent world average reads as follows 
\cite{oldeman}:
\begin{equation}\label{DMs-bound}
\Delta M_s > 16.6 \, \mbox{ps}^{-1} \, \mbox{(90\% C.L.)},
\end{equation}
and is already close to the SM expectation. 

In contrast to the situation in the $B_d$-meson system, the width difference
$\Delta\Gamma_s$ between the mass eigenstates is expected to be sizeable.
This quantitiy may provide interesting studies of CP violation through 
``untagged'' $B_s$ rates \cite{dun,FD-CP}, which are defined as 
\begin{equation}
\langle\Gamma(B_s(t)\to f)\rangle
\equiv\Gamma(B^0_s(t)\to f)+\Gamma(\bar B^0_s(t)\to f),
\end{equation}
and are characterized by the feature that we do not distinguish between
initially, i.e.\ at time $t=0$, present $B^0_s$ or $\bar B^0_s$ mesons. 
In such untagged rates, the rapidly oscillating $\Delta M_s t$ terms cancel.
Although $B$-decay experiments at hadron 
colliders should be able to resolve the $B^0_s$--$\bar B^0_s$ oscillations, 
untagged $B_s$ rates are interesting in terms of efficiency, 
acceptance and purity. Recently, the first results for $\Delta\Gamma_s$ 
were reported from the Tevatron, using the $B^0_s\to J/\psi\phi$ channel \cite{DDF}:
\begin{equation}
\frac{|\Delta\Gamma_s|}{\Gamma_s}=\left\{
\begin{array}{ll}
0.65^{+0.25}_{-0.33}\pm0.01 & \mbox{(CDF \cite{CDF-DG})}\\
0.24^{+0.28+0.03}_{-0.38-0.04} & \mbox{(D0 \cite{D0-DG})}.
\end{array}
\right.
\end{equation}
It will be interesting to follow the evolution of the data for this quantity. 
Let us next have a look at important benchmark decays of $B_s$ mesons.
For a more detailed recent discussion, see Ref.~\cite{RF-landscape}.

\subsection{CP Violation in $B^0_s\to J/\psi \phi$}
The decay $B^0_s\to J/\psi \phi$ is the counterpart of $B^0_d\to J/\psi K_{\rm S}$, 
where we have just to replace the down spectator
quark by a strange quark. In contrast to the $B_d$ case, the final state of
$B^0_s\to J/\psi \phi$ consists of two vector mesons. Consequently, we have to
deal with an admixture of different CP eigenstates in the final state. However,
these can be disentangled through the angular distribution of the
$B^0_s\to J/\psi[\to\ell^+\ell^-] \phi [\to K^+K^-]$ decay products \cite{DDLR}. 
The corresponding
observables show essentially negligible direct CP violation in the SM, and allow
the determination of $\sin\phi_s$ through mixing-induced CP violation \cite{DDF}.
In the SM, we have $\phi_s=-2\lambda^2\eta={\cal O}(10^{-2})$, so that 
we expect here a {\it tiny} value of $\sin\phi_s$, i.e.\ {\it tiny} mixing-induced
CP violation in $B^0_s\to J/\psi \phi$. Needless to note, the big hope is that
experiments will find a {\it sizeable} value of $\sin\phi_s$, which would give us
an immediate signal for CP-violating NP contributions to $B^0_s$--$\bar B^0_s$ 
mixing \cite{NiSi}--\cite{DFN}. For specific scenarios of NP where such effects
may actually arise, see, for instance, Refs.~\cite{JN}--\cite{Z-prime}.

\subsection{$B_s\to D_s^\pm K^\mp$ and $B_d\to D^\pm \pi^\mp$}
The decays $B_s\to D_s^\pm K^\mp$ \cite{BsDsK} and $B_d\to D^\pm \pi^\mp$
\cite{BdDpi} can be treated on the same theoretical basis, and provide new strategies 
to determine the UT angle $\gamma$ \cite{RF-gam-ca}. These modes are pure ``tree" 
decays, and can generically be written as $B_q\to D_q \bar u_q$. 
Looking at the corresponding decay topologies, it can be seen that a
characteristic feature of these modes is that both a $B^0_q$ and a $\bar B^0_q$ 
meson may decay into the same final state $D_q \bar u_q$. Consequently,  
interference effects between $B^0_q$--$\bar B^0_q$ mixing and decay processes 
emerge, which allow us to probe the weak phase $\phi_q+\gamma$ through 
measurements of the corresponding time-dependent
decay rates. 

In the case of $q=s$, i.e.\ $D_s\in\{D_s^+, D_s^{\ast+}, ...\}$ and 
$u_s\in\{K^+, K^{\ast+}, ...\}$, these interference effects are governed 
by a hadronic parameter $X_s e^{i\delta_s}\propto R_b\approx0.4$, where
$R_b\propto |V_{ub}/V_{cb}|$ is the usual side of the UT, and hence are large. 
On the other hand, for $q=d$, i.e.\ $D_d\in\{D^+, D^{\ast+}, ...\}$ 
and $u_d\in\{\pi^+, \rho^+, ...\}$, the interference effects are described 
by $X_d e^{i\delta_d}\propto -\lambda^2R_b\approx-0.02$, and hence are tiny. 

The observables provided by the $\cos(\Delta M_qt)$ and $\sin(\Delta M_qt)$ 
terms of the time-dependent decay rates allow a {\it theoretically clean} 
determination of $\phi_q+\gamma$ \cite{BsDsK,BdDpi}.
Since $\phi_q$ can be determined separately, $\gamma$ can be extracted.
However, there are also problems in this approach. First, we encounter an 
{\it eightfold} discrete ambiguity for $\phi_q+\gamma$, which reduces the power
for the search of NP effects considerably. Second, in the case of $q=d$, an 
additional input is required to extract $X_d$ since interference effects proportional
to $X_d^2={\cal O}(0.0004)$ would otherwise have to be resolved, which is not possible
from a practical point of view. 

A combined analysis of the $B_s^0\to D_s^{(\ast)+}K^-$ and 
$B_d^0\to D^{(\ast)+}\pi^-$ modes allows us to solve these problems 
\cite{RF-gam-ca}. Since these decays are related through the interchange
of all down and strange quarks, the $U$-spin flavour symmetry of strong
interactions, which is an $SU(2)$ subgroup of the full $SU(3)_{\rm F}$, 
provides an interesting playground. Following these lines, 
an {\it unambiguous} value of $\gamma$ can be extracted from the 
corresponding observables. To this end, $X_d$ has {\it not} to be fixed, and $X_s$ 
may {\it only} enter through a $1+X_s^2$ correction, which is determined through
untagged $B_s$ rates. First studies were recently performed for the LHCb
experiment, and look very promising \cite{wilkinson-CKM}.

\subsection{The $B_s\to K^+K^-$, $B_d\to\pi^+\pi^-$ System}
The $B^0_s\to K^+K^-$ decay is a $\bar b \to \bar s$ transition, and
involves tree and penguin amplitudes, as the $B^0_d\to\pi^+\pi^-$ mode 
\cite{RF-BsKK}. However, because of the different CKM structure, the latter 
topologies play actually the dominant r\^ole in the $B^0_s\to K^+K^-$ channel. 
In analogy to (\ref{Adirpi+pi-}) and (\ref{Amixpi+pi-}), its CP asymmetries can 
be written in the following generic form:
\begin{eqnarray}
{\cal A}_{\rm CP}^{\rm dir}(B_s\to K^+K^-)&=&
G_1'(d',\theta';\gamma) \label{CP-BsKK-dir-gen}\\
{\cal A}_{\rm CP}^{\rm mix}(B_s\to K^+K^-)&=&
G_2'(d',\theta';\gamma,\phi_s),\label{CP-BsKK-mix-gen}
\end{eqnarray}
where $(d',\theta')$ are the counterparts of $(d,\theta)$. Since $\phi_s$ is 
negligibly small in the SM -- or can be determined through $B^0_s\to J/\psi \phi$ 
should CP-violating NP contributions to $B^0_s$--$\bar B^0_s$ mixing make it 
sizeable -- we may convert the measured values of these observables into
a theoretically clean contour in the $\gamma$--$d'$ plane. In a similar manner,
the CP asymmetries of $B^0_d\to\pi^+\pi^-$ can be converted into a theoretically
clean contour in the $\gamma$--$d$ plane. A key feature of the 
$B^0_s\to K^+K^-$ and $B^0_d\to\pi^+\pi^-$ decays is that they are related
to each other through an interchange of all down and strange quarks, 
and that there is a one-to-one correspondence between the decay topologies. 
Consequently, the $U$-spin flavour symmetry implies 
the following relations:
\begin{equation}\label{U-spin-rel}
d'=d, \quad \theta'=\theta.
\end{equation}
Applying the former, we may extract $\gamma$ and $d$ through the 
intersections of the $\gamma$--$d$ and $\gamma$--$d'$ 
contours \cite{RF-BsKK}. Moreover, we may determine $\theta$ and $\theta'$, 
which allow an interesting internal consistency check of the second $U$-spin 
relation in (\ref{U-spin-rel}). Detailed experimental feasibility studies show that
this strategy is very promising for LHCb \cite{LHCb-analyses}, allowing
an experimental accuracy for $\gamma$ of just a few degrees. Concerning
possible $U$-spin-breaking corrections, the relations in (\ref{U-spin-rel}) 
are particularly robust as they involve only ratios of hadronic amplitudes,
where all $SU(3)$-breaking decay constants and form factors cancel in 
factorization and also chirally enhanced terms would not lead to corrections. 
Moreover, the determination of $\theta$ and $\theta'$ offers an internal consistency
check, as we have noted above, and the contours in the $d$--$\theta$ and
$d$--$\theta'$ planes allow a very transparent discussion of $U$-spin-breaking 
effects. In the numerical examples discussed in Ref.~\cite{RF-BsKK}, and most 
recently in Ref.~\cite{RF-landscape}, taking the newest data for the 
$B\to\pi\pi,\pi K$ system into account, the situation would be remarkably stable with 
respect to even large $U$-spin-breaking corrections to (\ref{U-spin-rel}), which 
appear not very likely.

\subsection{$B^0_s\to\mu^+\mu^-$}
In the SM, this rare decay and its conterpart $B^0_d\to\mu^+\mu^-$
originate from $Z^0$-penguin and box diagrams and are strongly suppressed. 
Since the matrix elements of the corresponding low-energy effective Hamiltonian 
involve only the decay constants $f_{B_q}$ of the $B_q$ mesons, we 
encounter a very favourable situation with respect to hadronic effects. 
The SM branching ratios
can then be written in the following compact form \cite{Brev01}:
\begin{eqnarray}
\lefteqn{\mbox{BR}( B_s \to \mu^+ \mu^-) = 4.1 \times 10^{-9}}\nonumber\\
&&\qquad\qquad\times \left[\frac{f_{B_s}}{0.24 \, \mbox{GeV}} \right]^2 \left[
\frac{|V_{ts}|}{0.040} \right]^2 \left[
\frac{\tau_{B_s}}{1.5 \, \mbox{ps}} \right] \left[ \frac{m_t}{167 
\, \mbox{GeV} } \right]^{3.12}\label{BR-Bsmumu}\\
\lefteqn{\mbox{BR}(B_d \to \mu^+ \mu^-) = 1.1 \times 10^{-10}}\nonumber\\
&&\qquad\qquad\times\left[ \frac{f_{B_d}}{0.20 \, \mbox{GeV}} \right]^2 \left[
\frac{|V_{td}|}{0.008} \right]^2
\left[ \frac{\tau_{B_d}}{1.5 \, \mbox{ps}} \right] \left[
\frac{m_t}{167 \, \mbox{GeV} } \right]^{3.12}.\label{BR-Bdmumu}
\end{eqnarray}
The simultaneous measurement of the $B_d\to\mu^+\mu^-$ and 
$B_s\to\mu^+\mu^-$ branching ratios would allow a determination of
the UT side $R_t$ that is complementary to that provided by 
$\Delta M_d/\Delta M_s$. Moreover, also correlations between the 
$B_q\to \mu^+\mu^-$ branching ratios and the mass differences $\Delta M_q$ 
can be derived in models with ``minimal flavour violation", which include also
the SM, that are more robust with respect to $SU(3)$-breaking 
effects \cite{Buras-rel}.

The most recent upper bounds from CDF are given as follows
\cite{CDF-Bmumu}:
\begin{equation}\label{Bmumu-exp}
\mbox{BR}(B_s\to\mu^+\mu^-)<1.5\times10^{-7}, \quad
\mbox{BR}(B_d\to\mu^+\mu^-)<3.9 \times10^{-8},
\end{equation}
and are still about two orders of magnitude away from the SM. 
Consequently, should the $B_q \to \mu^+ \mu^-$ decays 
be governed by their SM contributions, we could only 
hope to observe them at the LHC. On the other hand, 
since the $B_q \to \mu^+ \mu^-$ transitions originate from 
FCNC processes, they are sensitive probes of NP. In particular, 
the branching ratios may be dramatically enhanced in specific NP (SUSY) 
scenarios, as was recently reviewed in Ref.~\cite{buras-NP}. Should this 
actually be the case, these decays could already be seen at run II of the 
Tevatron, and the $e^+e^-$ $B$ factories could observe $B_d\to \mu^+ \mu^-$. 
Let us finally emphasize that the current experimental bounds on 
$B_s\to\mu^+\mu^-$ can also be converted into bounds on NP parameters
in specific scenarios. In the context of the constrained minimal 
supersymmetric extension of the SM (CMSSM) with universal scalar masses, 
such constraints were recently critically  discussed by the authors of Ref.~\cite{EOS}.

\section{Conclusions and Outlook}\label{sec:concl}
Thanks to the efforts at the $B$ factories, CP violation is now well established
in the $B$-meson system. The exploration of this phenomenon is characterized
by a fruitful interplay between theory and experiment. The data have shown
that large non-factorizable hadronic effects arise in non-leptonic $B$ decays, 
so that the challenge for a reliable theoretical description within dynamical QCD approaches remains, despite interesting recent progress. Concerning weak 
interactions, the Kobayashi--Maskawa mechanism of CP violation has successfully 
passed its first experimental tests, in particular through the comparison between the 
measurement of $\sin 2\beta$ through $B^0_d\to J/\psi K_{\rm S}$ and the 
CKM fits. However, the most recent average for $(\sin2\beta)_{\psi K_{\rm S}}$ is 
somewhat on the lower side, and there are a couple of puzzles in the $B$-factory 
data. It will be very interesting to monitor these effects, which could be first hints for 
physics beyond the SM, as the data improve. Moreover, it is crucial to refine the 
corresponding theoretical analyses further and to explore correlations with other 
flavour probes. 

Despite this impressive progress, still many aspect of $B$ physics have not
yet been studied. For instance, $b\to d$ penguin processes are now entering the 
stage,  since lower SM bounds for the corresponding branching ratios 
are found to be very close to the corresponding experimental upper limits. 
Moreover, also the $B_s$-meson system, which cannot be studied
with the BaBar and Belle experiments, is still essentially unexplored
and plays an outstanding r\^ole for the further testing of the flavour sector of the 
SM. After new results from run II of the Tevatron, the promising $B_s$ physics 
potential can be fully exploited at the LHC, in particular by LHCb. Moreover,
it is important to complement the $B$-decay studies through other flavour probes,
where rare $K\to\pi\nu\bar\nu$ decays are particularly interesting.

With the exception of a couple of flavour puzzles, which do not yet allow us to draw 
definite conclusions, the SM is still in good shape. However, the observed 
neutrino oscillations as well as the evidence for dark 
matter and the baryon asymmetry of the Universe tell us that the SM is 
incomplete. Moreover, specific NP scenarios contain usually also new 
sources of flavour and CP violation, which may manifest themselves at the 
flavour factories. In the autumn of 2007, also the LHC is expected 
to go into operation, which will provide insights  into electroweak symmetry 
breaking and, hopefully, also give us direct evidence for NP through the 
production and subsequent decays of new particles in the ATLAS and CMS 
detectors. Obviously, there should be a very fruitful interplay between these 
``direct" NP studies and the ``indirect" information provided by the 
flavour-physics sector that is  currently addressed in detail within a 
new workshop \cite{workshop}. In view of these promising perspectives,
an exciting future should be ahead of us!

\vspace*{0.3truecm}

\noindent
{\bf Acknowledgements}\\
\noindent
I would like to thank the organizers for inviting me
to this very enjoyable event. On this occasion, I would like to 
congratulate Gustavo once again to his birthday and would like
to express my best wishes.

\vspace*{-0.3truecm}

\end{document}